\documentclass[12pt,preprint]{aastex}
\newcommand \lsol{L$_{\odot}$}
\newcommand \msol{M$_{\odot}$}

\newfont{\rten}{cmr10}

\begin{document}

%\normalsize

\title{The dust, planetesimals and planets of HD 38529}

\author{Amaya Moro-Mart\'{\i}n\altaffilmark{1}, 
Renu Malhotra\altaffilmark{2}, 
John M. Carpenter\altaffilmark{3}, 
Lynne A. Hillenbrand\altaffilmark{3}, 
Sebastian Wolf\altaffilmark{4}, 
Michael R. Meyer\altaffilmark{5}, 
David Hollenbach\altaffilmark{6},
Joan Najita\altaffilmark{7} and
Thomas Henning\altaffilmark{4}}

\altaffiltext{1}{Department of Astrophysical Sciences, Peyton Hall, Ivy Lane, 
Princeton University, Princeton, NJ 08544, USA, amaya@astro.princeton.edu}

\altaffiltext{2}{Department of Planetary Sciences, University of Arizona, 1629 E. 
University Boulevard, Tucson, AZ 85721, USA}

\altaffiltext{3}{Department of Astronomy, California Institute of Technology, 
Pasadena, CA 91125, USA}

\altaffiltext{4}{Max-Planck-Institut f\"ur Astronomie, K\"onigstuhl 17, 69117 
Heidelberg, Germany} 

\altaffiltext{5}{Steward Observatory, University of Arizona, 933 North Cherry Ave., 
Tucson, AZ 85721, USA}

\altaffiltext{6}{NASA Ames, Moffet Field, CA 94035, USA}

\altaffiltext{7}{National Optical Astronomy Observatory, 950 North Cherry Ave.,  
Tucson AZ 85721, USA} 

\begin{abstract}
HD~38529 is a post-main sequence G8III/IV star (3.5 Gyr old) with a planetary system 
consisting of at least two planets having $\it{M}$sin$\it{i}$ of 
0.8~M$_{Jup}$ and 12.2~M$_{Jup}$, semimajor axes of 
0.13 AU and 3.74 AU, and eccentricities of 0.25 and 0.35, respectively.
$\it{Spitzer}$ observations show that HD~38529 has an excess emission above 
the stellar photosphere, with a signal-to-noise ratio (S/N) at 70 $\mu$m of 4.7, 
a small excess at 33 $\mu$m (S/N=2.6) and no excess $<$30 $\mu$m. 
We discuss the distribution of the potential dust-producing planetesimals
from the study of the dynamical perturbations of the two known planets, considering
in particular the effect of secular resonances. We identify three dynamically stable 
niches at 0.4--0.8 AU, 20--50 AU and beyond 60 AU. We model the spectral energy 
distribution of HD~38529 to find out which of these niches show signs 
of harboring dust-producing plantesimals. 
The secular analysis, together with the SED modeling resuls, 
suggest that the planetesimals responsible for most of the dust emission 
are likely located within 20--50 AU, a configuration that resembles that of the Jovian planets + 
Kuiper Belt in our Solar System. Finally, we place upper limits 
(8$\times$10$^{-6}$~lunar masses~of 10 $\mu$m particles) to the amount of dust 
that could be located in the dynamically stable region that exists between the 
two planets (0.25--0.75 AU).
\end{abstract}

\keywords{circumstellar matter --- Kuiper Belt
--- infrared: stars
--- planetary systems
--- stars: HD 38529
} 

\section{Introduction}
 
HD~38529 is a post-main sequence star (G8 III/IV) with an estimated age
of 3.5$\pm$1 Gyr (Valenti \& Fischer~\citeyear{vale05} and Gonzalez et al.~\citeyear{gonz01}), 
a distance of 42$\pm$2 pc (Perryman et al.~\citeyear{perr97}), and 
T$_{eff}$ = 5697 K, L$_{*}$ = 6.31~\lsol, M$_{*}$ = 1.47~\msol~
[Fe/H] = 0.445 (Valenti \& Fischer~\citeyear{vale05}). 
High precision radial velocity monitoring of HD~38529 has led to the discovery 
of a close-in Jupiter-mass planet (HD~38529b -- Fischer et al.~\citeyear{fisc01}) 
and a second more massive and more distant planet (HD~38529c -- 
Fischer et al.~\citeyear{fisc03}). Butler et al. (\citeyear{butl06})
report the current estimates of the parameters of the two planets: 
masses ($\it{M}$sin$\it{i}$) of 0.8~M$_{Jup}$ and 12.2~M$_{Jup}$, 
semimajor axes of 0.13 AU and 3.74 AU, 
and eccentricities of 0.25 and 0.35, for HD~38529b and  HD~38529c respectively. 

As part of the $\it{Spitzer}$ Legacy Program FEPS (Meyer et al.~\citeyear{meye06}), 
we searched for debris disks around the nine stars in the FEPS sample known from 
radial velocity (RV) studies to have one or more massive planets
(Moro-Mart\'{\i}n et al.~\citeyear{ama07}). 
HD~38529 was found to be the only star in that sub-sample to have an excess emission above the stellar 
photosphere, with a signal-to-noise ratio (S/N) at 70 $\mu$m of 4.7, a 
small excess at 33 $\mu$m (S/N = 2.6) 
and no excess at $\lambda$ $<$ 30 $\mu$m (Moro-Mart\'{\i}n et al.~\citeyear{ama07}). 
HD~38529 therefore joined the small group of stars known to date that have both IR 
excess and one or more known planetary companions. Table 1 summarizes the properties of these systems, 
showing a wide diversity of planetary architectures. 
Six of these nine sources, HD 33636, HD 50554, HD 52265, HD 82943, HD 117176 
and HD 128311, are similar to HD~38529 in that their
$\it{Spitzer}$ observations also show an excess at 70 $\mu$m~but 
no excess at 24 $\mu$m, 
implying that the bulk of the excess emission is arising from cool material (T$<$100 K) 
located mainly beyond 10 AU (Beichman et al.~\citeyear{beic05a}). This means that
these stars not only harbor planets (as inferred from their radial velocity observations) 
but also harbor an outer belt of dust-producing planetesimals (responsible for 
their IR excess) and in this regard they resemble the Solar System in its
Jovian planets +  Kuiper Belt configuration. The other two sources in Table 1 (HD 69830
and $\epsilon$-Eri) have warm dust and are therefore less relevant to the present discussion. 

In this paper we constrain the distribution of the potential dust-producing planetesimals
from the study of the dynamical perturbations of the two known planets, considering
in particular the effect of secular resonances. This allows us to identify three regions
where planetesimals could be dynamically stable ($\S$2). After examining the lifetimes 
of the dust grains and concluding that the dust traces the location of the parent 
planetesimals ($\S$3), we show how we can further constrain the location of the dust (and the 
parent planetesimals) by modeling HD 38529 spectral energy distribution (SED -- $\S$4). 
In $\S$5 we discuss together the results from the dynamical and SED analysis, and finally $\S$6
summarizes our results. 

\section{Possible Location of the Dust-Producing Planetesimals: Effect of Gravitational Perturbations by the Planets}

We can identify the possible location of the dust-producing planetesimals by studying
the effect of the planetary perturbations on the stability of the planetesimals' orbits. 
Consider the orbital parameter space of semimajor axis and eccentricity, $(a,e)$. 
First, we can eliminate planetesimals with orbits that would 
cross the orbits of the planets, i.e., orbits that have 
either periastron or apoastron within the periastron-apoastron range
of each of the planets.  Second, we note that planetesimals in initially circular 
orbits would be strongly unstable in the vicinity of the orbits 
of each of the two known planets, in a range of semimajor axis, 
$\Delta a \simeq \pm 1.5 (m_{\hbox{planet}}/m_\star)^{2/7}$ 
(Duncan et al.~\citeyear{dunc89}).
These two considerations identify several regions where planetesimals could not
be stable, shown as the grey and red shaded zones in Fig.~1. In addition to these
perturbations that operate over short timescales, we need to consider the effect of 
perturbations operating over much longer timescales. These are the secular perturbations, 
and as we will see below, they can cause a strong eccentricity excitation of the planetesimals, 
particularly at secular resonance locations, that can significantly shorten their lifetime.

We now calculate the effect of secular perturbations of the two planets on the 
planetesimals (taken as test particles on initially circular orbits). 
For the two planets, we assume co-planar orbits and minimum masses, taking
their orbital parameters from Butler et al. (\citeyear{butl06}).
For a two planet system, there are two linear modes that excite eccentricities
of test particles; we follow the Laplace-Lagrange secular perturbation analysis 
(see Murray and Dermott~\citeyear{murr99}).  The frequencies, $g_i$, phases, $\beta_i$, and 
amplitudes, $E^{(i)}$, of the two secular modes for the HD 38529 system are:
\begin{equation}
{g_1 = 0.106 \hbox{arcsec/yr}, E^{(1)} = (6.697\times10^{-3}, 0.3300), \beta_1=0.2271}
\end{equation}
\begin{equation}
{g_2 = 19.7 \hbox{arcsec/yr}, E^{(2)} = (0.2787, -6.411\times10^{-5}), \beta_2=1.594}
\end{equation}

The secular variations of the eccentricity vectors of planets b,c are 
given by

\begin{equation}
{e_b\cos\varpi_b = E^{(1)}_1\cos(g_1t+\beta_1)+E^{(2)}_1\cos(g_2t+\beta_2)}
\end{equation}
\begin{equation}
{e_b\sin\varpi_b = E^{(1)}_1\sin(g_1t+\beta_1)+E^{(2)}_1\sin(g_2t+\beta_2)}
\end{equation}
\begin{equation}
{e_c\cos\varpi_c = E^{(1)}_2\cos(g_1t+\beta_1)+E^{(2)}_2\cos(g_2t+\beta_2)}
\end{equation}
\begin{equation}
{e_c\sin\varpi_c = E^{(1)}_2\sin(g_1t+\beta_1)+E^{(2)}_2\sin(g_2t+\beta_2)}
\end{equation}

The fast mode dominates in planet b's eccentricity vector,
whereas the slow mode dominates planet c's eccentricity vector.

Now consider a test particle (i.e. a planetesimal) in this system.
Its secular perturbations have two components: 
a ``free'' apsidal precession due to the overall quadrupole potential 
(arising from the combined effects of the post-Newtonian stellar 
gravitational field, and the gravitational effects of planetary masses 
approximated by uniform density circular rings of radius equal to their 
semimajor axes), and a ``forced'' eccentricity-apsidal perturbation due to 
the planetary secular modes (i.e., due to the secular variations of the 
elliptical planetary orbits).  The amplitude of the latter perturbation 
is a function of the semimajor axis of the test particle; at some values 
of semimajor axis, where the free precession frequency is close to a 
planetary secular frequency, the test particle is subjected to a large 
amplitude resonant eccentricity excitation. To determine the maximum 
forced eccentricity obtained by a test particle, we use the secular 
resonance analysis given in Malhotra (\citeyear{malh98}) which includes the effects 
of non-linear saturation of the amplitude of the resonantly forced 
eccentricity from each secular mode.
The results are shown in Fig.~1: a test particle initially in a circular 
orbit would have its eccentricity excited by the slow mode to the values
indicated by the blue curve; the green curve is for the eccentricity
excited by the fast mode.  

Clearly, very significant secular eccentricity excitation occurs over 
a wide zone that extends to distances much larger than the unstable zones 
identified above. Remarkably, even though the strongly unstable zone of the outer 
planet extends outward to only about 5.5 AU, the eccentricity excitation exceeds 
0.3 to more than 10 AU.  In general, we see that the forced eccentricities exceed 0.1 
everywhere up to $\sim57$ AU.  A strong secular resonance with 
the slow mode occurs at a semimajor axis value of $\sim55$ AU, where the 
particles become planet-crossing or can even collide with the star;
from the secular analysis (Malhotra 1998), we calculate that the timescale 
to excite the eccentricity from $\sim0$ to $\sim1$ is $\sim1.9\times10^8$ years.
(An analogous phenomenon occurs at the inner edge of the asteroid belt 
in our Solar System, at the location of the so-called $\nu_6$ secular 
resonance where asteroidal apsidal precession rates are resonant with the 
6th secular mode of the Solar System.)  Thus, tentatively, we can
identify three regions that could harbor planetesimal populations
(or perhaps small planets) in low eccentricity orbits: 
an inner region 0.4--0.8 AU in-between the two planets' orbits, and two outer 
regions, 20--50 AU and beyond 60 AU. This conclusion is confirmed by numerical 
integrations of test particles as demonstrated in Figures 2 and 3. 

What does the high eccentricity excitation of test particles imply
for dust-producing planetesimal populations? 
Because these are forced eccentricities, neighboring particle orbits 
could have a high degree of apsidal alignment, so that the relative
velocities of planetesimals would be determined by Kepler shear, and
therefore would be small. However, it is likely that, in the formation
and evolution process, free eccentricities of planetesimal populations 
become comparable to the forced eccentricities, as observed in the solar 
system's asteroid belt and Kuiper Belt.  In that case, the forced
eccentricities would indeed be a measure of the relative velocities 
for collisions amongst planetesimals in these regions.
Thus, if the zones identified above are populated by planetesimals,
the planetary perturbations would cause mutual planetesimal collisions 
that would result in the production of dust. 

Alternatively, because the primordial disk of HD 38529 was massive enough to form 
at least two giant planets, arguably, these zones could harbor one or more large 
planetesimals, or even sub-Jovian mass planets so far undetected: 
sufficiently massive planets would tend to suppress the secularly forced eccentricities, and 
clear out planetesimal populations from their vicinities, 
thereby removing sources of dust; while large 1000 km-size planetesimals 
could have stirred up and ground away the planetesimal disk at early times. 

\section{A Collision-Dominated System: the Dust Traces the Planetesimals}

To evaluate whether the HD~38529 debris disk is radiation-dominated or collision-dominated we
consider the lifetimes of the dust grains: 
\begin{itemize}

\item The collisional lifetime at a distance $\it{R}$ from the star can be 
estimated as t$_{col}$ = P$_{orb}$/8$\Sigma$$\sigma_{geo}$ (Backman \& Paresce~\citeyear{back93}), 
where P$_{orb}$ is the orbital period, $\Sigma$ is the surface density of the 
grains and $\sigma_{geo}$ is the grain geometric cross section. 
 $\Sigma$$\sigma_{geo}$ is the dimensionless fractional surface area of the disk, of
the order of L$_{dust}$/L$_*$ $\sim$ 10$^{-5}$(5600/T$_*$)$^3$(F$_{70,dust}$/F$_{70,*}$)
(Backman \& Paresce~\citeyear{back93}). 
For HD~38529,  F$_{70,*}$ = 17.4 mJy, F$_{70,dust}$ = 66.9 mJy, T$_*$ = 5697 K 
and M$_{*}$ = 1.47 \msol~ (Moro-Mart\'{\i}n et al.~\citeyear{ama07}), so that 
L$_{dust}$/L$_{*}$ = 3.6 $\times$10$^{-5}$ and 
\begin{equation}
{t_{col}  = 3.4 \times 10^3 P_{orb} = 2.8  \times 10^3 ({R \over AU})^{3/2} yr.}
\end{equation}

\item The  Poynting-Robertson (P-R) lifetime at distance $\it{R}$, i.e. the time it takes for 
the dust grain to migrate from $\it{R}$ to the star, is given by 
\begin{equation}
{t_{PR} = {4 \pi b \rho \over 3} {c^2 R^2 \over L_*} = 710 
({b \over \mu m}) ({\rho \over g/cm^3}) ({R \over AU})^2 
({L_\odot \over L_*}) {1 \over 1+albedo} ~yr}
\end{equation}
(Burns, Lamy \& Soter,~\citeyear{burn79}; Backman \& Paresce~\citeyear{back93}), 
where $\it{b}$ and $\rho$ are the grain radius and density, respectively.  
More relevant is the time it takes for the particle to drift under
P-R drag from a populated to a relatively unpopulated region, i.e. the time it would
take to fill out the gap under P-R drag ($\it{t_{PR}^{fill}}$). 
If the scale over which the dust density 
significantly decreases is $\it{x}$\% of $\it{R}$, this time is
\begin{equation}
{t_{PR}^{fill} =  (1-(1-{x \over 100})^{2}) ~t_{PR} = 2.0 \times 10^4 
({R \over AU})^2 yr,}
\end{equation}
where we have assumed that x = 10\%, L$_{*}$ = 6.3 \lsol, $\rho$ = 2.5 g/cm$^3$, 
$\it{b}$ = 10 $\mu$m and $\it{albedo}$ = 0.1. 

\end{itemize}

If the sources of dust are outside the orbit of the planet, and if P-R drag dominates
the dynamics, as the dust particles drift inward due to P-R drag they 
are likely to be scattered out of the system when crossing the orbit of a planet, 
creating a dust depleted region inside the planet's orbit
(Roques et al.~\citeyear{roqu94}; Liou \& Zook~\citeyear{liou99}; 
Moro-Mart\'{\i}n \& Malhotra~\citeyear{ama02},~\citeyear{ama03}). 
The ejection is very efficient for planets in circular orbits: 
planets with masses of 3--10 M$_{Jup}$ located 
between 1--30 AU eject $>$90\% of the particles that go past their orbits, while a 
1 M$_{Jup}$ planet at 30 AU ejects $>$ 80\% of the particles, and about 50\%--90\%
if located at 1 AU (Moro-Mart\'{\i}n \& Malhotra~\citeyear{ama05b}); 
these results are valid for dust particle sizes in the range
0.7--135~$\mu$m (assuming astronomical silicate composition), and when the central
star is solar-type. 

On the other hand, if the system is collision-dominated (Krivov et al.~\citeyear{kriv00}; 
Dominik \& Decin~\citeyear{domi03}; Wyatt~\citeyear{wyat05}), 
mutual collisions can fragment the larger particles to smaller and smaller 
sizes, until they are blown out from the system by radiation pressure. This means 
that the dust particles would be destroyed before they migrate far from the 
location of their parent bodies under P-R drag, i.e. the dust traces the location of 
the parent bodies. In this case, an inner cavity in 
the dust density distribution could only arise if the planetesimals themselves 
are confined to a belt, i.e. if there is an inner edge to their spatial distribution. 
This scenario may suggest a massive planet confining the inner edge of the 
dust-producing planetesimals. 

For the HD~38529 system, the estimates in equations [7] 
and [9] suggest that $\it{t_{col}}$ $<<$ $\it{t_{PR}^{fill}}$ at all relevant radii, 
i.e., we are in the collision-dominated regime. This means that the dust 
emission traces the location of the dust-producing planetesimals, with 
dynamically stable niches at 0.4--0.8 AU, 20--50 AU and beyond 60 AU. To find out  
which of these niches do actually show signs of harboring dust-producing plantesimals 
we turn now to the study of the IR excess emission detected by $\it{Spitzer}$.

\section{SED Modeling}

\subsection{Single Temperature Models are Insufficient}

The 70 $\mu$m observations of HD~38529 are not spatially resolved and therefore
it is not possible to know unambiguously the spatial distribution of the dust.
We can learn about the characteristic temperature of the dust from the spectral 
energy distribution (Hillenbrand et al. in preparation). 
From the ratio of the excess emission at 70 $\mu$m to that at 33 $\mu$m, and 
assuming these wavelengths are at the Wien's tail of the dust blackbody emission, 
the ratio of the excess fluxes yield a characteristic temperature of of 43$\pm$4 K.
If the blackbody grains are in thermal equilibrium, we can calculate the 
location of the dust from
\begin{equation}
{{R_{dust} \over AU} = ({L_* \over L_{\sun}})^{1/2}({278 K \over T_{dust}})^2}
\end{equation}
(Backman et al.~\citeyear{back93}); for  HD~38529, R$_{dust}$ = 106$\pm$18 AU. The 
dust mass can then be estimated by assuming the dust is in a thin shell at 
a distance R$_{dust}$, and that the grains are spherical with cross section
equal to their geometric cross section, so that the total number of dust particles is 
$\sim$(4$\pi$R$_{dust}^2$/$\pi$$\it{b}^2$)$\times$$\tau$ and 
M$_{dust}$ $\sim$ 16$\pi$R$_{dust}$$^2$$\tau$$\it{b}$$\rho$/3. The optical 
depth, $\tau$, can be approximated as L$_{dust}$/L$_{*}$ so that, 
\begin{equation}
{{M_{dust} \over M_{\earth}} \sim 6.28\times10^{-5}({L_{dust} \over L_*})({\rho \over g cm^{-3}})({<b> \over \mu m})({R_{dust} \over AU})^2}
\end{equation}
(Jura et al.~\citeyear{jura95}), where $\it{b}$ and $\rho$ are the particle's 
radius and density. 
It is important to note that the above estimate is a lower limit: in the absence of 
sensitive (sub)millimeter detections, no realistic constraints 
can be made to the dust mass, a significant fraction of which could be locked in grains with
sizes of $\sim$1 mm that emit efficiently in the (sub)millimeter but contribute little to 
the infrared emission. For HD~38529, L$_{dust}$/L$_{*}$ = 3.6$\times$10$^{-5}$, and
using  $\it{b}$ = 10~$\mu$m and $\rho$ = 2.5 g/cm$^{-3}$ we obtain 
M$_{dust}$ $>$ 1.9$\times$10$^{-9}$~\msol\footnote{Dust mass estimates for the KB dust disk range 
from a total dust 
mass $<$ 3$\times$10$^{-10}$~\msol~(Backman et al.~\citeyear{back95}) to 
$\sim$ 4$\times$10$^{-11}$~\msol~for dust particles $<$ 150 $\mu$m 
(Moro-Mart\'{\i}n \& Malhotra~\citeyear{ama03}); with a 
fractional luminosity of L$_{dust}$/L$_*$$\sim$10$^{-7}$--10$^{-6}$ (Stern~\citeyear{ster96}). 
The fractional luminosity of the asteroid belt dust (a.k.a zodiacal cloud) is estimated to 
be L$_{dust}$/L$_*$$\sim$10$^{-8}$--10$^{-7}$ (Dermott et al.~\citeyear{derm02}).}.

Another estimate of the dust temperature can be made from fitting a photosphere plus a single 
temperature blackbody to the IRS spectrum only; this results in a dust temperature of 79 K, which
would corresponds to a distance of 31 AU. The discrepancy with the characteristic temperature 
derived from the ratio of the excess emission at 70 $\mu$m to that at 33 $\mu$m indicates 
that it is not possible to conclude that the dust is confined to a narrow ring as implied by 
the assumption of a single grain temperature. Similarly, Hillenbrand et al. (in preparation) 
find that the SEDs of about 1/3 of the FEPS targets with 70~$\mu$m excess emission are better 
fit by multi-temperature rather than single-temperature blackbody models.
The need for a multi-temperature grain distribution has previously been 
found in systems like $\beta$-Pic (e.g. Li \& Greenberg~\citeyear{li98} and Telesco 
et al.~\citeyear{tele05}) and HR4796A (e.g. Li \& Lunine~\citeyear{li03a}), 
and has been unambiguously confirmed by the 
eight spatially resolved observations in scattered light of debris disks known to date, 
which led Kalas et al. (\citeyear{kala06}) to conclude that 
debris disks show two basic architectures: 1) belts about 20--30 AU 
wide and with well-defined outer boundaries (HR 4796A, Fomalhaut and HD 139664); 
and 2) wider belts with sensitivity-limited edges implying widths $>$50 AU 
(HD 32297, $\beta$-Pic, AU-Mic, HD 107146 and HD 53143).
In the absence of spatially resolved observations, the SED modeling of HD~38529 
(or any other source under consideration) should enable to explore a range of 
disk architectures.

\subsection{Multi Temperature (SED) Models}
\subsubsection{SED Modeling Assumptions} 

For the modeling of the SED, we use the radiative transfer code developed by 
Wolf \& Hillenbrand (\citeyear{wolf03}). Because of the above considerations, 
we model the dust disk as an annulus of inner radius R$_{in}$, outer
radius R$_{out}$, total dust mass M$_{dust}$,
and a constant surface density ($\Sigma \propto r^0$, so that the number density, 
n(r)~$\propto$~r$^{-1}$).
We assume that the dust grains are composed of silicates\footnote{For a study on how the SEDs depend on the grain composition we refer to Wolf \& Hillenbrand 
(\citeyear{wolf03}) and Moro-Mart\'{\i}n, Wolf \& Malhotra (\citeyear{ama05a}).} with 
optical constants from Weingartner \& Draine (\citeyear{wein01}). 
For the particle sizes we consider two options:
1) a single grain size of 10 $\mu$m in radius, and 2) a particle size distribution following
a power law, $\it{n(b)}$ $\propto$ $\it{b^{-q}}$, where $\it{b}$ is the particle radius, 
$\it{q}$ = 3.5 (for grains in collisional equilibrium), 
$\it{b_{min}}$ = 2 $\mu$m and 
$\it{b_{max}}$ = 10 $\mu$m.  In both cases, the radius of 10 $\mu$m was chosen 
because such a grain radiates efficiently at 70 $\mu$m.
Because of the last property, this choice of grain size provides a
lower limit for M$_{dust}$. The value selected for $\it{b_{min}}$ corresponds to the 
``blow-out'' size, i.e. the minimum size of the grains that can remain bound in the 
system, given by
\begin{equation}
{ {b_{min} \over \mu m} = 0.52 \times {2.5g/cm^{-3} \over \rho} \times {1 + albedo \over 1.1}
\times {L_*/L_{\odot} \over (M_*/M_{\odot})(T_*/5780K)}}
\end{equation}
(Artymowicz et al.~\citeyear{arty89}). For HD~38529, L$_{*}$ = 6.31 \lsol, M$_{*}$ = 1.47 \msol~and  
T$_{*}$ = 5697 K; using a grain density $\rho$ = 2.5 g/cm$^{-3}$ 
and albedo = 0.1, we get $\it{b_{min}}$ = 2 $\mu$m . 

We assume that the dust disk is optically thin with the dust being in thermal 
equilibrium with the stellar radiation field. 
Only scattering, absorption and reemission of stellar radiation by dust 
grains are taken into account, neglecting scattering and dust heating 
from the infrared radiation produced by the optically thin dust disk. 
With these assumptions, the temperature of the dust grain,
for a given dust size and a chemical composition, depends only on the distance to the 
central star
\footnote{Our SED modeling assumes that the observed dust excess arises
from a cicumstellar disk. However, the outermost planet has a $\it{M}$sin$\it{i}$ of 12.2~M$_{Jup}$, 
placing it in the boundary between planets and brown dwarfs. Even though the latter 
could potentially harbor a disk, the prevalence of cold dust in the HD 38529 system indicates
that this could not be the dominant source of the observed dust because this disk 
would be located near the star (the outermost planet's semimajor axis is 3.74 AU)
implying the presence of warm dust that is not observed.}.

The outer radius of the disk, R$_{out}$, cannot be constrained with data currently available. 
%without multiwavelength sub-mm and mm photometry or spatially resolved observations.
Based on scattered-light observations from 
nearby debris disks, disk sizes of about 100 AU 
to several hundred AU are inferred (Kalas et al.~\citeyear{kala05}; Dent et al.~\citeyear{dent00}; 
Greaves et al.~\citeyear{grea00}; Wilner et al.~\citeyear{wiln02}; 
Holland et al.~\citeyear{holl03}; Liu~\citeyear{liu05}; Metchev et al.~\citeyear{metc05}; 
Ardila et al.~\citeyear{ardi04}). 
%However, submillimeter observations of the 
%center of the Trapezium (Williams, Andrews \& Wilner~\citeyear{will05}) 
%and of the Taurus region indicate 
%that the majority of primordial disks may be smaller (in particular if they
%are born within a cluster). 
For this reason we consider disk sizes of R$_{out}$ = 50 AU (Solar System size), 
R$_{out}$ = 100 AU and R$_{out}$ = 500 AU.

With the above three values for R$_{out}$, assuming a uniform density distribution and
with the grain size and composition fixed, we then vary R$_{in}$ and
M$_{dust}$ (our only two free parameters) to create a grid of models where
we allow R$_{in}$ to vary from the silicate sublimation radius (R$_{sub}$, 
where T$_{sub}$ = 1550 K) to R$_{out}$. 
This accounts for the possibility of having either a dust disk of wide 
radial extent or a narrow ring of dust. 

The observations to be modeled are the $\it{Spitzer}$-IRS\footnote{The IRS spectrum for $\lambda$ $<$ 14.21 $\mu$m 
(short-low module) shows a small offset with respect to the Kurucz model. Because the emission at those
wavelengths is clearly photospheric, we have multiplied the spectrum for 
$\lambda$ $<$ 14.21 $\mu$m by 1.045 to make it 
coincide with the stellar photosphere (so these wavelengths do not dominate the $\chi$$^2$ 
statistics). In addition, the IRS spectrum shows a small discontinuity at 14.21 $\mu$m, 
possibly because the star was not centered in the slit of the long-low module. To correct
for this discontinuity, we have multiplied the IRS spectrum for $\lambda$ $>$ 14.21 $\mu$m 
by 1.108. Accounting for this, the corrected 33 $\mu$m flux is 95$\pm$2 mJy, where the 
error is the internal uncertainty. For the calculation of the S/N of the excess 
we used the total uncertainty, 
obtained from adding in quadrature the internal uncertainty and the calibration uncertainty, 
the latter taken to be 6\%. This is an underestimate of the significance of the departure from 
a pure photosphere because this calibration uncertainty is for the overall spectrum, not 
for individual wavelengths relative to each other.} synthetic 
photometric points at 13 $\mu$m, 24 $\mu$m  and 33 $\mu$m and the $\it{Spitzer}$-MIPS
photometric points at 24 $\mu$m and 70 $\mu$m (from Moro-Mart\'{\i}n et al.~\citeyear{ama07}). 
To evaluate whether a particular model is a valid fit to these data points, or if not, to what 
degree of certainty the model can be excluded, we calculate its $\chi$$^{2}$ probability, 
P($\chi$$^{2}$ $\mid$ $\nu$), where $\nu$ is the number of degrees of freedom. In 
our case $\nu$ = 2, as the only two free parameters are R$_{in}$ and M$_{dust}$, and
all the other disk parameters and dust properties are fixed to the values given above. 
%This is the probability that the observed $\chi$$^2$ for the correct model is less than 
%the value of $\chi$$^{2}$ for the model under consideration, with P(0 $\mid$ $\nu$) = 0 
%and  P(1 $\mid$ $\nu$) = 1. 
The probability is defined so that P(0 $\mid$ $\nu$) = 0 and  P(1 $\mid$ $\nu$) = 1. 
Models with P($\chi$$^{2}$ $\mid$ $\nu$) $>$ 0.9973 can be 
excluded with a 3-$\sigma$ certainty, while models with P($\chi$$^{2}$ $\mid$ $\nu$) $>$ 0.683 
could be excluded with a 1-$\sigma$ certainty. 

\subsubsection{SED Modeling Results}

Figure 4 shows the two-dimensional grids of SED models described above: 
in $\it{green}$ are the models with 
P($\chi$$^{2}$ $\mid$ $\nu$) $<$ 0.683, i.e. models that are consistent with
the observations; $\it{red}$ represents models with P($\chi$$^{2}$ $\mid$ $\nu$) $>$ 0.683,
i.e. models that could be excluded with a 1-$\sigma$ certainty; and $\it{black}$
corresponds to models with  P($\chi$$^{2}$ $\mid$ $\nu$) $>$ 0.9973, i.e. models
that can be excluded with a 3-$\sigma$ certainty. 
Some of these SEDs are shown in Fig. 5 together with the observations. 
As can be seen in both figures, the models are degenerate and for a given R$_{out}$, 
there are many pairs of R$_{in}$ and M$_{dust}$ that could fit the observed SED. 
Because of this degeneracy, our main interest lies in identifying the models that can 
be excluded. These are of particular interest because they allow us to identify dust depleted regions
whose origin can be studied in terms of the overall dynamics of the planetary system (see $\S$2).

The main results from Fig. 4(a--c) regarding the models that assume 
grains 10 $\mu$m in radius are the following
\begin{itemize}
\item For R$_{out}$ = 50 AU, we can exclude models 
with R$_{in}$ $\la$ 5 AU with a certainty of 3-$\sigma$, or
we could exclude models with R$_{in}$ $\la$ 14 AU with a certainty of 
1-$\sigma$; 
the data is consistent with models having 15 AU $\la$ R$_{in}$ $\la$ 50 AU,
i.e., models ranging from a 35 AU wide disk (extending from 15 AU--50 AU) 
to a very  narrow ring of dust located at 50 AU. 
In other words, the observations require the presence of an inner cavity 
in the dust density distribution of at least 5 AU in radius, with larger 
cavities consistent with the data (the latter implying narrower rings
of larger dust mass). 
\item For R$_{out}$ = 100 AU, none of the R$_{in}$ considered 
(ranging from the dust sublimation radius, R$_{sub}$, to 100 AU) can be
excluded with a 3-$\sigma$ certainty; while models with R$_{in}$ $\la$ 7 AU
and those with R$_{in}$ $\ga$ 40 AU could be excluded to 1-$\sigma$. Possible
fits include models with 8 AU $\la$ R$_{in}$ $\la$ 35 AU.
\item For R$_{out}$ = 500 AU, again none of the R$_{in}$ considered 
(from R$_{sub}$ to 400 AU) can be excluded with a 3-$\sigma$ certainty; 
while models with R$_{in}$ $\ga$ 14 AU
could be excluded to 1-$\sigma$. Possible models 
include those with  R$_{sub}$ $\la$ R$_{in}$ $\la$ 12 AU. 
\end{itemize}

We can understand these results as follows. For a constant surface density,
$\Sigma \propto r^0$, there is 100 times more dust mass between 50 and 100 AU 
than there is between 5 and 10 AU, while the dust temperature is proportional 
to $\it{r}$$^{-1/2}$. This results in that there is a significant contribution
to the 70 $\mu$m emission from the largest radii, R$_{out}$. To keep the 
70 $\mu$m emission constant, the dust surface area or mass at R$_{out}$ must be similar 
as R$_{out}$ varies, i.e. $\Sigma$R$_{out}$ $\sim$ constant, this means that: 
1) larger R$_{out}$ requires
a smaller $\Sigma$, and smaller $\Sigma$ means that we can tolerate smaller inner 
radii since there is now less dust mass placed there; this explains 
why the SEDs can be fitted with a smaller inner cavity for 
R$_{out}$ = 100 AU, and with a constant surface density all the way to the sublimation 
radius in the 500 AU case; and 2) smaller R$_{out}$ requires a larger $\Sigma$,
and in order to keep the 24 $\mu$m and 33 $\mu$m fluxes below their upper limits, this
requires to increase the inner cavity size. A larger inner cavity would also be needed if 
we had assumed $\Sigma \propto r^{-1}$ instead of constant. 

In the models shown in Fig. 4(d) we relax the assumption that 
all the grains are 10 $\mu$m in radius and allow for the presence of 
smaller grains, with the particle size distribution following
$\it{n(b)}$ $\propto$ $\it{b^{-3.5}}$, and with $\it{b_{min}}$ = 2 $\mu$m and 
$\it{b_{max}}$ = 10 $\mu$m. We find that most of the models can be 
excluded with a 3-$\sigma$ certainty, i.e. we can exclude the presence of 
a significant population of small grains inside 100 AU based on the lack of a significant 
continuum emission at $\lambda$ $<$ 30 $\mu$m. A similar conclusion has been obtained from 
several other debris disks whose spectroscopy show little or no solid state
features, indicating that the dust grains have sizes $\ga$ 10 $\mu$m
(e.g. Jura et al.~\citeyear{jura04}; Stapelfeldt et al.~\citeyear{stap04}).

\section{Discussion of the Dynamical and the SED Analysis}

\subsection{Location of Cold Dust}

The two outermost regions identified in $\S$2 where planetesimals could 
survive for extended periods of time in the presence of the two known radial 
velocity-detected planets, namely 20--50 AU and beyond 60 AU, are in broad agreement 
with the allowed dust locations that result from the modeling of the SED.
In particular, we saw that the models with R$_{out}$ = 50 AU predicted 
an inner cavity of radius $>$ 5 AU;  while a range of dust disks with R$_{out}$ = 50 AU 
and R$_{in}$ $>$ 20 AU could fit the observed SED. From the models with 
R$_{out}$ = 100 AU, we can conclude that it is very unlikely that 
the observed dust emission arises only from planetesimals located beyond 60 AU because
the models with R$_{out}$ = 100 AU and R$_{in}$ = 60 AU, 70 AU, 80 AU and 90 AU 
have P($\chi$$^{2}$ $\mid$ $\nu$) larger than 0.83, 0.87, 0.89 and 0.91, respectively 
A similar conclusion can be drawn from the models with R$_{out}$ = 500 AU, 
and R$_{in}$ $>$ 60 AU, for which P($\chi$$^{2}$ $\mid$ $\nu$) $>$ 0.94.
Therefore, from the dynamical and SED modeling we conclude that the planetesimals
responsible for most of the dust emission are likely located within the 
20--50 AU region. 

Given that the maximum eccentricities in the 20--50 AU region are moderately high, 
0.25--0.3 (see Figures 1 and 2) -- implying collisional velocities at 30 AU of approximately 
0.3$\times$$\it{v_{circ}}$ = 2.8 km/s -- 
could a moderate rate of dust production persist in this relatively old system?
(A ``moderate rate" is important in order for the dust production to be long-lived, 
instead of being depleted rapidly in the early history of the system). 
Following Wyatt et al. (\citeyear{wyat07}), we can estimate 
the maximum fractional luminosity of the excess, $\it{f_{max}}$, that could originate from 
a planetesimal belt of a given age that is evolving in quasi-steady state. 
Wyatt's analytical model assumes that the 
planetesimals and the dust are in collisional equilibrium, and that their size distribution follows
a continuous power law of index -3.5. For HD 38529, and given the results above, we assume that 
the planetesimal belt extends from 20--50 AU, with a mean planetesimal eccentricity 
of 0.3. Other parameters in the model are the diameter of the largest planetesimal in the 
cascade, 2000 km, and the specific incident energy required to catastrophically
destroy a planetesimal, 200 J/kg. At the age of the HD 38529 system, 3.5 Gyr, the model predicts
$\it{f_{max}}$ = 2.26$\times$10$^{-6}$ (see eq. [20] in Wyatt et al.~\citeyear{wyat07}). 
This value is smaller than the observed fractional 
luminosity of the excess, $\it{f_{obs}}$ = 3.6$\times$10$^{-5}$~(see $\S$3). However, because
there are two orders of magnitude uncertainty in the estimate of $\it{f_{max}}$, 
we cannot reject a scenario in which the dust observed is the result of the steady grinding 
down of planetesimals. If we were to assume that the above value of $\it{f_{max}}$ is correct, 
an estimate of the timescale over which the fractional luminosity can be mantained at the value of
$\it{f_{obs}}$ is given by $\it{t_{age}}$$\times$$\it{f_{max}}$/$\it{f_{obs}}$ = 220 Myr. 
In this case, a possible scenario could be that the stable region beyond 60 AU supplies some 
planetesimals that drift into the 20--50 AU region by non-gravitational effects. 

\subsection{Upper Limits on Warm Dust}

In $\S$2 we identified a  modestly stable small zone in-between the two known planets
(0.4--0.8 AU), that has also been identified in numerical
simulations by Barnes \& Raymond (\citeyear{barn04}); our secular perturbation
analysis provides a theoretical explanation for those numerical
stability results.  Raymond \& Barnes (\citeyear{raym06}) consider terrestrial planet 
accretion in this zone.  They conclude that HD 38529 is likely to support 
an asteroid belt and perhaps Mars-sized planets, but not larger planets, 
because the potential feeding zone for the accretion of a terrestrial 
planet would be limited by the high eccentricities of the planetesimals 
in this region. 
The lack of an IR excess at wavelengths shorter than 30 $\mu$m allows
us to place an upper limit on the amount of warm (asteroidal) dust that
could be located in this inner region.  We use the IRAC 5$\mu$m and 8$\mu$m,
IRS 13$\mu$m and MIPS 24$\mu$m photometric measurements, and we assume
10$\mu$m size silicate grains (optical constants taken from
Weingartner \& Draine~\citeyear{wein01}).  We find that a 3-$\sigma$ upper limit to the
dust mass in this potential asteroid belt is $3\times$10$^{-13}$ M$_\odot$ or 
10$^{-7}$~M$_{\earth}$. For comparison, the mass estimate for the zodiacal cloud in the 
terrestrial planet region of the Solar System is 3$\times$10$^{-10}$~M$_{\earth}$
(Hahn et al.~\citeyear{hahn02}), i.e. 330 times smaller than the estimated upper limit 
of warm dust in HD~38529.

\section{Conclusions}
HD~38529 harbors a planetary system consisting of at least two planets (with 
$\it{M}$sin$\it{i}$ of 0.8~M$_{Jup}$ and 12.2~M$_{Jup}$, semimajor axes of 0.13 AU 
and 3.74 AU, and eccentricities of 0.25 and 0.35) and a likely population of dust-producing 
planetesimals that are responsible for the 70 $\mu$m excess emission detected 
by $\it{Spitzer}$. Using analytical and numerical dynamical analysis, in this paper 
we have constrained the distribution of the potential dust-producing planetesimals 
from the study of the dynamical perturbations of the two known planets, considering 
in particular the effect of secular resonances. A dust disk inner edge at 5.5 AU 
would naturally arise from the gravitational scattering of planetesimals and dust 
grains by the outermost planet. We show that larger inner cavities in the dust disk, 
that would be consistent with the observed SED, can be created due to the secular effects 
that arise from the interaction between the two massive planets. From the analysis of 
the secular perturbations we identify three regions that could harbor planetesimal 
populations in low eccentricity orbits (where the planetesimals could be long-lived): 
0.4--0.8 AU, 20--50 AU and beyond 60 AU.  
From the modeling of the SED we conclude that the planetesimals responsible for most 
of the dust emission observed by $\it{Spitzer}$ are likely located within the 20--50 AU region.
In this regard, HD 38529 resembles the configuration of the Solar System's Jovian 
planets + Kuiper Belt (KB). 
The SED models give a dust mass estimate of 1--5$\times$10$^{-10}$~\msol~of 10 $\mu$m particles.
The presence of a significant population of small grains inside 100 AU
(with the particle size distribution following
$\it{n(b)}$ $\propto$ $\it{b^{-3.5}}$, and with $\it{b_{min}}$ = 2 $\mu$m and 
$\it{b_{max}}$ = 10 $\mu$m) is excluded to a 3-$\sigma$ certainty level based on the lack
of a significant continuum emission at $\lambda$ $<$ 30 $\mu$m. 
We do not find evidence of dust emission within the innermost region, 
with a 3-$\sigma$ dust mass upper limit of 10$^{-7}$~M$_{\earth}$ (in 10 $\mu$m
grains), suggesting any remnant dust belt would have a mass smaller than 330 times that in
the Solar System's zodiacal cloud. 

The SED models are degenerate. We need to break this degeneracy to get a better 
understanding of this planetary system, in particular, of how the spatial distribution
of the dust and the planetesimals are affected by the gravitational perturbations
of the two planets. This requires spatially resolved images to constrain the disk sizes, 
and/or high resolution spectroscopy observations to look for spectral features that 
could constrain the grain size and composition, and/or accurate photometric points 
in the 33 $\mu$m--70 $\mu$m range and in the sub-mm to better determine the shape of the SED. 

\begin{center} {\it Acknowledgments} \end{center}
We thank the rest of the FEPS team members, colleagues at the Spitzer Science Center, 
and members of all the Spitzer instrument teams for advice and support. 
This work is based [in part] on observations made with the Spitzer Space Telescope,
which is operated by the Jet Propulsion Laboratory, California Institute of Technology 
unders NASA contrast 1407. 
A.M.M. is under contract with the Jet Propulsion Laboratory (JPL) funded by NASA through
the Michelson Fellowship Program. JPL is managed for NASA by the California Institute of
Technology. A.M.M. is also supported by the Lyman Spitzer Fellowship at Princeton University.
R.M. and M.R.M. are supported in part through the LAPLACE node of NASA's Astrobiology Institute.
R.M. also acknowledges support from NASA-Origins of Solar Systems research program.
S.W. was supported through the DFG Emmy Noether grant WO 875/2-1 and WO875/2-2.
FEPS is pleased to acknowledge support from NASA contracts 1224768 and 1224566
administered through JPL.

\clearpage

%\begin{deluxetable}{lcccccc}
\begin{deluxetable}{llllllllllc}
\rotate
%\tablewidth{80pc}
\tablewidth{0pc}
\tablecaption{Stars with debris disks and planets\tablenotemark{a}}
\tablehead{
\colhead{Source} & 
%\colhead{Planet} &
\colhead{} &
\colhead{$\it{M}$sin$\it{i}$} &
\colhead{Period} &
\colhead{$\it{a}$} &
\colhead{$\it{e}$} &
\colhead{$\it{T_p}$} &
\colhead{$\omega$} &
\colhead{N$_{obs}$} &
\colhead{Excess\tablenotemark{b}} &
\colhead{Ref.}\\
\colhead{(HD $\#$)} & 
%\colhead{} &
\colhead{} &
\colhead{(M$_{Jup}$)} &
\colhead{(days)} &
\colhead{(AU)} &
\colhead{} &
\colhead{(JD-2440000)} &
\colhead{(deg)} &
\colhead{} &
\colhead{$\lambda$ [L$_{dust}$/L$_{*}$]} &
\colhead{}}
\startdata

 38529   & b    & 0.852           & 14.3093(13)   & 0.131      & 0.248(23)   & 9991.59(23) & 91.2(6.2) & 162 & 70$\mu$m [3.6x10$^{-5}$] & (1)\\
         & c    & 13.2            & 2165(14)      & 3.74       & 0.3506(85)  & 10085(15)   & 15.7(1.9) & 162 &  & \\
 33636   & b    & 9.28            & 2127.7(8.2)   & 3.27       & 0.4805(60)  & 11205.8(6.4)& 339.5(1.4)& 38  & 70$\mu$m [4.9$\times$10$^{-5}$] & (2)\\
 50554   & b    & 4.38            & 1223(12)      & 2.28       & 0.437(38)   & 10649(16)   & 7.9(4.3)  & 47  & 70$\mu$m [4.4$\times$10$^{-5}$] & (2)\\
 52265   & b    & 1.09            & 119.290(86)   & 0.504      & 0.325(65)   & 10833.7(4.2)& 243(15)   & 28  & 70$\mu$m [2.9$\times$10$^{-5}$] & (2)\\
 82943   & b    & 1.81            & 219.50(13)    & 0.752      & 0.39(26)    & \nodata     & 121.0(3.1)& 165 & 70$\mu$m [1.2$\times$10$^{-4}$] & (2)\\ 
         & c    & 1.74            & 439.2(1.8)    & 1.19       & 0.020(98)   & \nodata     & 260.0(10) & 165 & & \\
 117176  & b    & 7.49            & 116.6884(44)  & 0.484      & 0.4007(35)  & 7239.82(21) & 358.71(54)& 74  & 70$\mu$m [1.0$\times$10$^{-5}$] & (2)\\
 128311  & b    & 2.19            & 458.6(6.8)    & 1.10       & 0.25(10)    & 10211(76)   & 111(36)   & \nodata& 70$\mu$m [3.0$\times$10$^{-5}$] & (2)\\
         & c    & 3.22	          & 928(18)       & 1.76       & 0.170(90)   & 10010(400)  & 200(150)  & \nodata & & \\
 69830   & b    & 0.0322          & 8.6670(30)	  & 0.0789     & 0.100(40)   & 13496.800(60)&340(26)   & 74  & 24$\mu$m [2$\times$10$^{-4}$] & (3)\\
         & c    & 0.0374          & 31.560(40)	  & 0.187      & 0.130(60)   & 13469.6(2.8)& 221(35)   & 74 & & \\
         & d    & 0.0573          & 197.0(3.0)    & 0.633      & 0.070(70)   & 13358(34)   & 224(61)   & 74 & & \\
 $\epsilon$-Eri & b & 1.06        & 2500(350)     &  3.38      & 0.25(23)    & 8940(520)   & 6.15      & 120 & IR to sub-mm   &(4)\\
\tablenotetext{a}{Orbital parameters from Butler et al. (2006). $\it{a}$ and $\it{e}$ are the 
semimajor axis and eccentricity of the planet;  $\it{T_p}$ is the time of periastron passage (as a 
Julian day) and $\omega$ is the longitude of periastron. N$_{obs}$ is the number of radial velocity 
observations.  The number in parenthesis indicate the uncertainty in the last significant figures.}
\tablenotetext{b} {Wavelength at which the excess is detected, with the fractional luminosity of 
the excess in brakets. For stars with 70 $\mu$m excess, and assuming that the peak of the emission
is at 70 $\mu$m (T$_{dust}$ = 52.7 K), L$_{dust}$/L$_{\star}$ $\sim$ 10$^{-5}$(5600/T$_*$)$^3$(F$_{70,dust}$/F$_{70,*}$), 
where F$_{70,dust}$ and F$_{70,*}$ are the dust excess and photospheric flux at 
70~$\mu$m and T$_*$ is the 
stellar temperature (Bryden et al. (\citeyear{bryd06}). 
For HD 69830, the fractional luminosity of the excess is calculated by integrating the excess and photospheric emission
beteween 7 and 35 $\mu$m.
References are: (1)  Moro-Mart\'{\i}n et al. (2007); 
(2) Beichman et al. (2005a); 
(3) Beichman et al. (2005b) and
(4) Greaves et al. (1998, 2005).}
\enddata
\end{deluxetable}

\clearpage

\begin{figure}
\epsscale{1.0}
\plotone{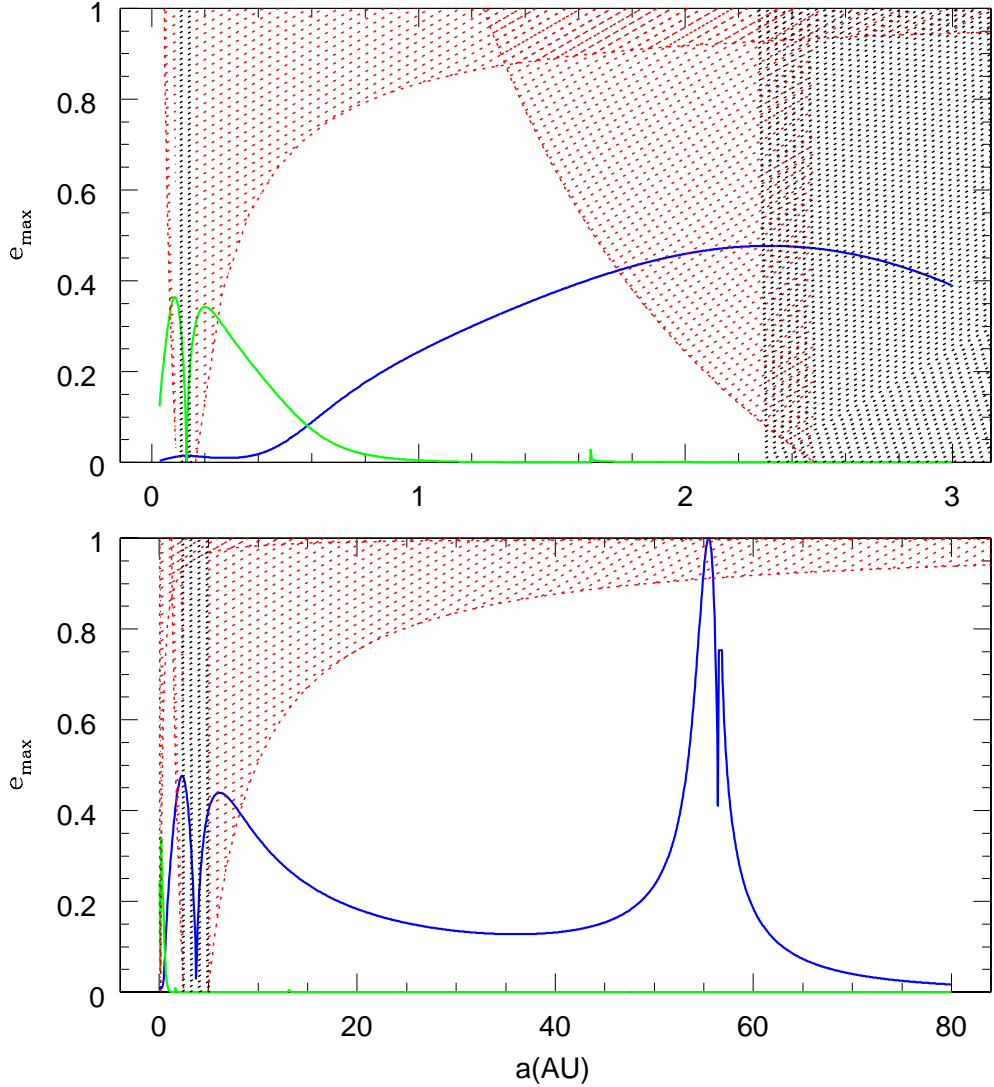}
\caption{Test particle orbits are planet-crossing (hence
unstable) in the red-shaded zone, and strongly unstable in the grey-shaded
zone (owing to overlapping first order mean motion resonances).  The secular
modes of the two planets, HD38529b and HD38529c, excite the eccentricities
of circular test particle orbits: the maximum eccentricity due to the fast
mode is shown by the green curve, while that due to the slow mode is shown
by the blue curve.  The inner region, interior to the outer planet,
is shown in the upper panel. The lower panel shows that the planetary
perturbations are very wide ranging: secular eccentricity excitation
exceeds 0.1 to nearly 60 AU; the sharp peak at 55 AU is due to a resonance
with the slow mode.}
\end{figure}

\clearpage

\begin{figure}
\epsscale{0.8}
\plotone{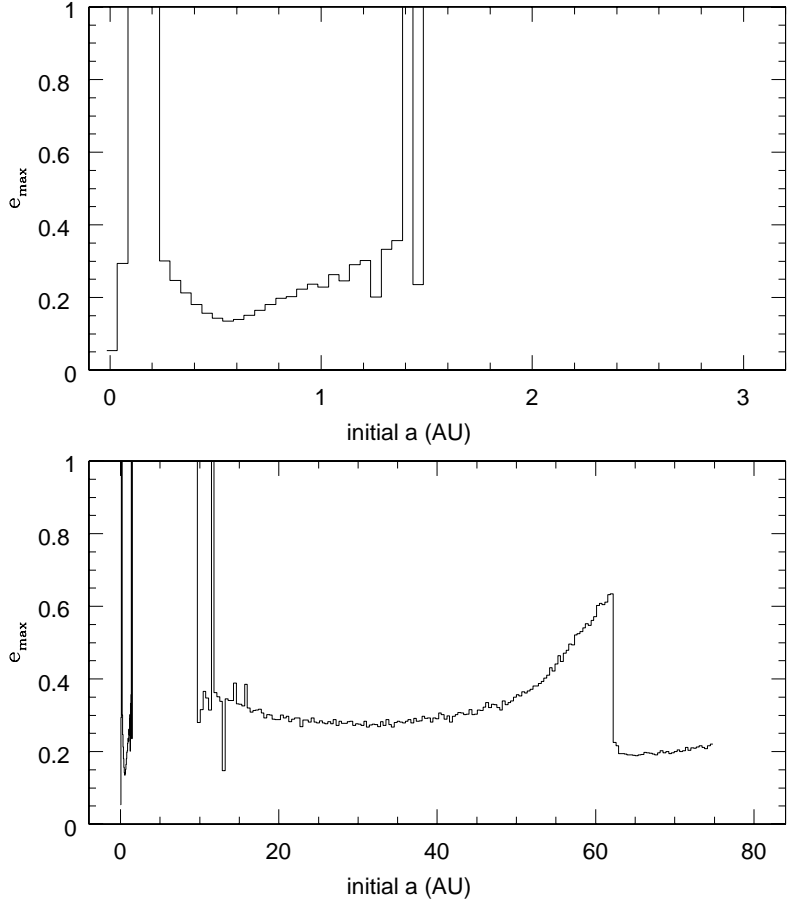}
\caption{Numerical simulations of 300 test particles in the HD 38529 system of two planets. The 
simulations include 100 particles uniformly spaced between 0.01--5 AU, 
and 200 particles uniformly spaced between 
5--75 AU, all in initially circular orbits coplanar with the planets.
The angular elements 
were chosen randomly between 0 and $2\pi$. 
Particles were removed if they approached the star closer
than 0.005 AU or approached a planet closer than the Hill
radius of the planet.
The orbits were integrated for 200 Myr using a symplectic 
integrator (Wisdom \& Holman~\citeyear{wisd91}). Overall, the simulations confirm the results from the secular analysis. 
They differ in that the maximum eccentricity in the 20--50 
AU region calculated from the secular analysis is smaller than that found in the numerical
integrations; this is because the latter include secular and non-secular perturbations (e.g. mean motion 
resonances), and the test particles have non-zero initial eccentricities and inclinations.}
\end{figure}

\clearpage

\begin{figure}
\epsscale{0.8}
\plotone{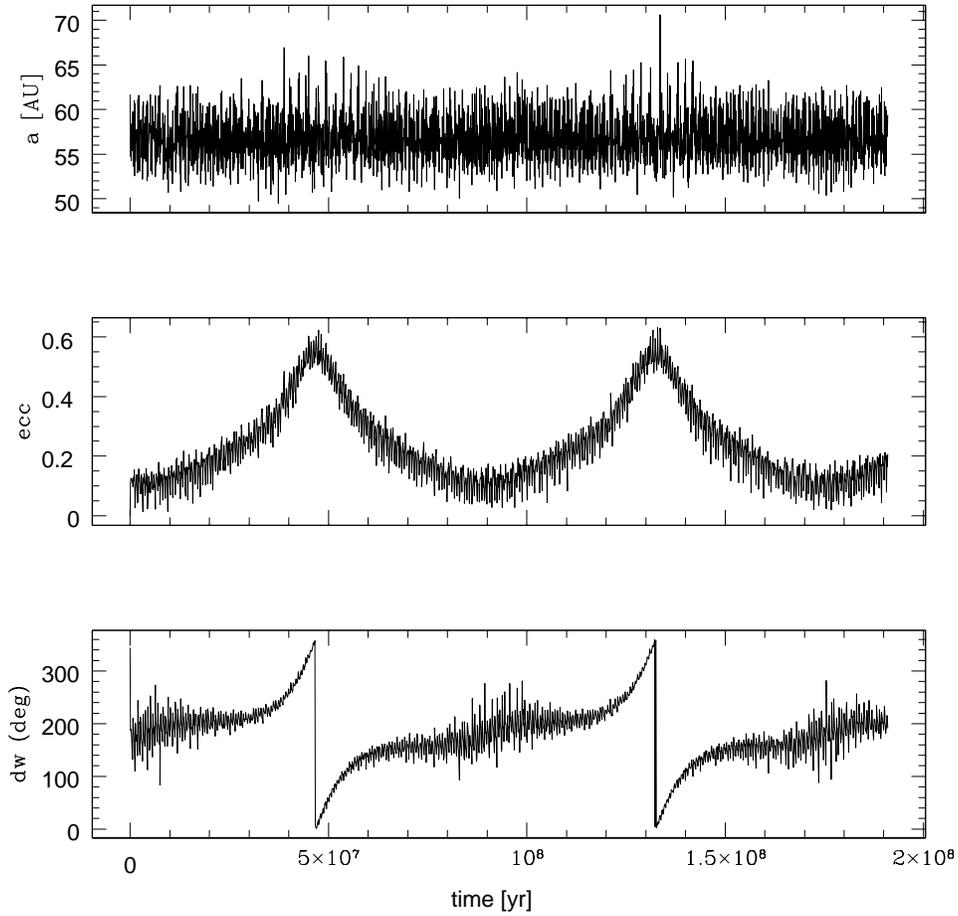}
\caption{Orbital evolution over the 200 Myr integration period of a test particle
near the secular resonance. The quantity  $\it{dw}$ is the difference between the
longitude of periastron of the particle and that of the outermost planet.
The timescale to excite the eccentricity is about 50 Myr.}
\end{figure}

\clearpage

\begin{figure}
\epsscale{0.8}
\plotone{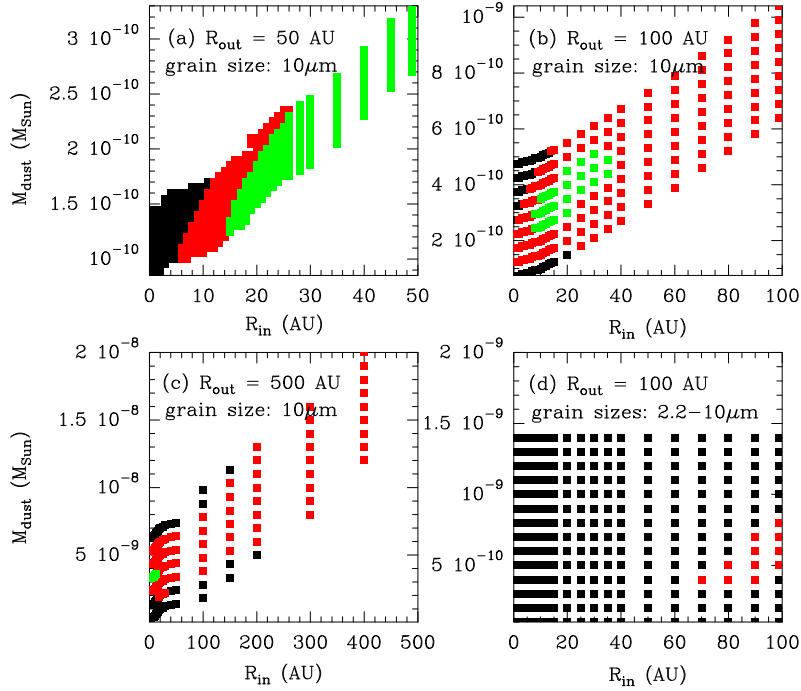}
\caption{Each point of these two dimensional grids represents a modeled SED, 
where R$_{in}$ and M$_{dust}$ are the two free parameters. Panels (a), (b) and (c) correspond to models 
with a single grain size of 10 $\mu$m and with R$_{out}$ = 50 AU, 100 AU 
and 500 AU, respectively. Panel (d) corresponds to models with a distribution of grain
sizes given by $\it{n(b)}$ $\propto$ $\it{b^{-3.5}}$, with $\it{b_{min}}$ = 2 $\mu$m and 
$\it{b_{max}}$ = 10 $\mu$m.  The models in $\it{green}$ are those with
P($\chi$$^{2}$ $\mid$ $\nu$) $<$ 0.683; $\it{red}$ in panels 
(a)--(c) represents models with 
P($\chi$$^{2}$ $\mid$ $\nu$) $>$ 0.683,
i.e. models that can be excluded with 1-$\sigma$ certainty; and $\it{black}$
corresponds to models with  P($\chi$$^{2}$ $\mid$ $\nu$) $>$ 0.9973, i.e. models
that are excluded with 3-$\sigma$ certainty. The red dots in panel (d) have 
P($\chi$$^{2}$ $\mid$ $\nu$) $>$ 0.988, i.e., they are all close to be excluded
with a 3-$\sigma$ certainty.}
\end{figure}

\clearpage

\begin{figure}
\epsscale{0.7}
\plotone{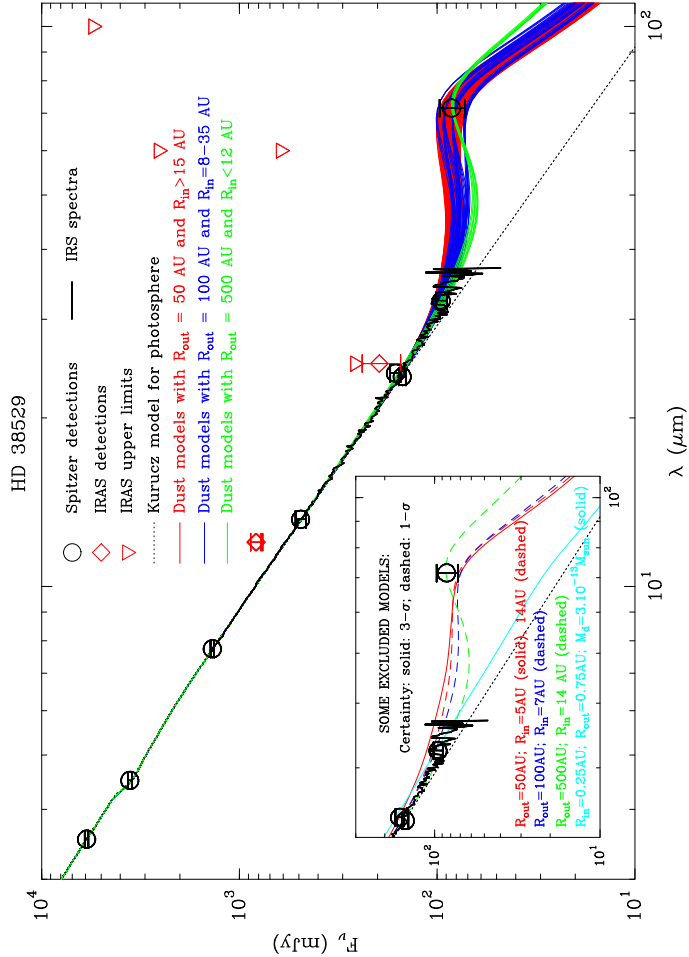}
\caption{Observed and modeled SEDs for HD~38529. 
The $\it{dotted~line}$ is the Kurucz model. The $\it{black~thick}$ 
line is the IRS low-resolution spectrum. The 
photometric points are identified as follows: $\it{black~circles}$ are the new Spitzer observations 
(IRAC, MIPS and synthetic photometry from IRS); $\it{red~diamonds}$ are IRAS observations. In all cases, 
the error bars correspond to 1-$\sigma$ uncertainties.  Upper limits are represented by $\it{triangles}$ 
and are given when F/$\Delta$F $<$ 3 and placed at F + 3$\times$$\Delta$F if F $>$ 0, or 3$\times$$\Delta$F 
if F  $<$ 0. The colored continuous lines in the main panel 
show three sets of models that fit the observations with a 
$\chi$$^{2}$ probability $<$ 0.68 (corresponding to the models represented in $\it{green}$ in
Fig. 4). The models include the emission from the photosphere and from a dust 
disk composed of astronomical silicate grains 10 $\mu$m in radius. 
We assume the dust disk extends from R$_{in}$ to R$_{out}$ with a constant surface density. 
We consider three values for R$_{out}$: 50 AU ($\it{red}$), 100 AU ($\it{blue}$) and 500 AU 
($\it{green}$). 
R$_{in}$ and M$_{dust}$ are allowed to vary.  The insert at the lower left shows 
the most relevant excluded models. The solid line represents models excluded
with a certainty of 3-$\sigma$, while the dashed line corresponds to 1-$\sigma$. The model
shown in $\it{light~blue}$ gives an upper limit to the amount of warm dust located between 0.25--0.75 AU.}

\end{figure}

\end{document}